# Two-stage WECC Composite Load Modeling: A Double Deep Q-Learning Networks Approach

Xinan Wang, *Student Member*, *IEEE*, Yishen Wang, *Member, IEEE*, Di Shi, *Senior Member, IEEE*, Jianhui Wang, *Senior Member, IEEE,* Zhiwei Wang, *Senior Member, IEEE*

*Abstract*—With the increasing complexity of modern power system, conventional dynamic load modeling with ZIP and induction motors (ZIP + IM) is no longer adequate to address the current load characteristic transitions. In recent years, the Western Electricity Coordinating Council Composite Load Model (WECC CLM) has shown to effectively capture the dynamic load responses over traditional load models in various stability studies and contingency analyses. However, a detailed WECC CLM model typically has a high degree of complexity, with over one hundred parameters, and no systematic approach to identifying and calibrating these parameters. Enabled by the wide deployment of PMUs and advanced deep learning algorithms, proposed here is a double deep Q-learning network (DDQN)-based, two-stage load modeling framework for the WECC CLM. This two-stage method decomposes the complicated WECC CLM for more efficient identification and does not require explicit model details. In the stage one, the DDQN agent determines a proper load composition that can approximate the true transient dynamics. In the second stage, the remaining parameters of the WECC CLM are selected with Monte-Carlo simulations. The proposed method shows that the identified load model is capable of accurately simulating the given dynamics of the reference load model. In addition, the identified load model has strong robustness to represent the reference load model under a wide range of contingencies. The proposed framework is verified using an IEEE 39-bus test system on commercial simulation platforms.

## I. Introduction

Accurate dynamic load modeling is critical for power system transient stability analysis and various simulation-based studies [1]-[2]. It is also known to improve the power system operation flexibility, reduce system operating costs, and better determine the corridor transfer limits [3]-[4]. In the past few decades, both industry and academic researchers have widely used ZIP and induction motors (ZIP + IM) as the composite load model (CLM) for quantifying load characteristics [5]-[7], in which ZIP approximates the static load transient behaviors and the IM approximates the dynamic load transient behaviors. This ZIP + IM load model has shown to be effective for simulating many dynamics in the power system, but in recent years, industry has started to observe various new load components, including single-phase IM, distributed energy resources (DER), and loads interfaced via power electronics that are being increasingly integrated into the system. The high penetration of these new types of loads brings profound changes to the transient characteristics at the load end, which raises the necessity for more advanced load modeling. For example, the well-known fault-induced, delayed-voltage-recovery (FIDVR) event is caused by the stalling of low-inertia single-phase IMs [8] when the fault voltage is lower than their stall thresholds. An FIDVR event poses potential voltage control losses and cascading failures in the power system [9]; however, FIDVR cannot be modeled by a conventional CLM model. Given these conditions, the Western Electricity Coordinating Council Composite Load Model (WECC CLM) is proposed.

To date, WECC CLM is available from multiple commercial simulation tools such as the DSATools$^{TM}$, GE PSLF, and PowerWorld Simulator. However, the detailed model structure, control logic, and parameter settings of the WECC CLM are limited by most of the software vendors (PowerWorld Simulator as a notable exception), and thus not transparent to the public [10], which impacts WECC CLM's general adoption and practicality. Furthermore, lack of detailed open-source information about the WECC CLM presents another major roadblock for conducting load modeling and parameter identification studies for system stability analysis.

Current WECC CLM works can be classified into two groups, which are component-based methods that rely on load surveys [11], [12] and measurement-based numerical fitting methods [13], [14]. In [11] and [12], the WECC CLM's parameters are estimated from surveys of different customer classes and load type statistics. However, the granularity and accuracy of the survey data depend entirely on the survey agency, and there are many assumptions being made that cannot be definitively verified. In addition, the survey is generally not up to date and does not reflect real-time conditions. In practice, all these limitations bring challenges in modeling the actual dynamic responses.

In another approach, authors in [13] and [14] numerically solve the parameter-fitting problem using nonlinear least squares estimators. In these methods, the parameter identifiability assessment and dimension reduction are conducted through sensitivity and dependency analysis. Though sensitivity analysis reflects the impacts of the individual parameter on the load dynamics, it fails to capture the mutual dependency between two or more parameters, which has been proved to be of great importance in composite load

This work is funded by SGCC Science and Technology Program under contract no. SGSDYT00FCJS1700676. X. Wang, Y. Wang, D. Shi and Z. Wang are with GEIRI North America, San Jose, CA 95134, USA. X. Wang and J. Wang are with the Department of Electrical and Computer Engineering, Southern Methodist University, Dallas, TX 75205 USA.

dynamics [15], [16]. In [14], the authors define the parameter dependency as the similarity of their influences on the dynamic response trajectory. Such a dependency analysis still falls short in factoring in the impact of multiple parameters on the load transient dynamics at the same time. In fact, with over one hundred parameters in the WECC CLM, the true interactions among them are hard to fully assess.

This paper proposes a double deep Q-learning network (DDQN)-based load modeling framework to conduct load modeling on the WECC CLM. In this framework, the DDQN agent serves as an optimization tool to estimate the key parameters in WECC CLM. Similar deep reinforcement learning-based optimization problems have also been discussed in [17] and [18]. In [17], the authors apply the deep deterministic policy gradient (DDPG) algorithm to determine the values of five control setpoints in the cooling systems to minimize the total data center cooling costs; in [18], the authors use the deep Q-learning network (DQN) to identify the runtime parameter in computer systems to improve the accuracy of the cache expiration time estimation. For this proposed load modeling work on WECC CLM, rather than directly constructing states for all parameters, the first stage only builds states for the load component fractions. These load component fractions or states serve as the "abstracted features" to characterize the full composite load models. The sequential decision making [19] then relates to assess whether these states can consistently represent the load model for future actions or load fraction changes. The DRL agent thus learns the Q-values for these states to obtain high rewards in the long run. In this way, the identified load component fractions are relatively robust to various conditions, which are another desirable property for load modeling. The remaining parameters, including the other top sensitive parameters as suggested in [14] and [20], are identified in the second stage.

This method adopts the Transient Security Assessment Tools (TSAT) from DSATools as the DDQN agent's training environment, which follows the state-of-art WECC model validation progresses to comply with industry practitioners. As such it is different from most nonlinear least square estimator-based load modeling work. The method recasts the load modeling for the WECC CLM into a two-stage learning problem. In the first stage, a DDQN agent is trained to find a load composition ratio that most likely represents the true dynamic responses at the bus of interest. Then, in the second stage, Monte-Carlo simulations are conducted to select the rest of the load parameters for the load model. From the Monte-Carlo simulations, the one set of parameters that best approximates the true dynamic responses is chosen for the load model. The specification [21] of the WECC CLM indicates that each load component in the model represents the aggregation of a specific type of load. Under such a composite load structure, it has been observed in [22] and [23] that different load composition ratios could have very similar transient dynamics. Therefore, solving the load composition ratio first and conducting the load parameter identification based on the identified ratio can significantly reduce the problem's complexity and increase load parameter identification computational efficiency. In addition, each parameter is independently selected in stage two through Monte-Carlo simulations, and the parameter identification criteria is to evaluate the dynamic response reconstruction. This method implicitly considers the dependency between two or more parameters. Our proposed method offers the following unique features and contributions:

1) *A load modeling framework for the WECC CLM with limited prior knowledge to model details.* Only the dynamic response curve is required to implement the proposed learning framework.
2) *The load model identified by this framework is robust to various contingencies.* The fitted load model is verified to be effective to recover the true dynamics with different fault locations, fault types and fault durations.
3) *The proposed method is scalable to different composite load structures*: In the DDQN training environment, the action taken by the agent is designed to be the load fraction changes on different load types. This set up allows the proposed method to be scaled from conventional CLM load models such as ZIP + IM to larger load models like the WECC CLM. The method can be easily extended when WECC CLM is updated with new load components, such as distributed energy resources (DERs).
4) *Applicable with limited data scenario*: Unlike other data-hungry supervised and unsupervised machine learning methods, our DDQN approach only needs a few sets of transient records to conduct load modeling, which effectively overcome the data availability issue.

The remainder of this paper is organized as follows. Section II discusses the load component definition in the WECC CLM and the associated parameter selection range of each component. Section III introduces the DDQN training environment formulation and the customized reward function. Section IV presents the case studies to validate the effectiveness of the proposed method using the DSATools$^{TM}$. Section V provides concluding remarks and discussions on future research.

## II. WECC CLM INTRODUCTION

### A. WECC CLM Structure

The WECC CLM is widely recognized as the state-of-the-art load model [24] due its robustness in modeling a variety of load compositions and its capability of simulating the electrical distance between the end-users and the transmission substations [9].

The detailed load structure for the WECC CLM is shown in Fig. 1, which mainly consists of three parts: substation, feeder, and load. The parameters for substation and feeder parts, such as the substation shunt capacitance $B_{ss}$ and transformer tap settings [25] usually follow the industry convention and do not have significant variance [23]-[27]. Therefore, in this paper, we set the feeder and substation parameters following industrial standard values [23]. The load in WECC CLM includes three three-phase induction motors, one single-phase induction motors, one electronic load, and one ZIP static load. Our load modeling work focuses on the load composition and parameter identification for these load components.

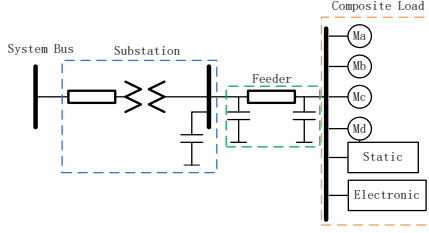

Fig. 1. WECC CLM structure [21].

### B. Three-phase Induction Motors

As shown in Fig 1, four motors are connected at the end-use bus. Three of them are three-phase induction motors, which are defined as **Ma**, **Mb**, and **Mc** in our system setup. **Ma**, **Mb** and **Mc** use the same fifth-order induction motor model shown from (1) - (9), which are derived from the three-phase motor model block diagram given by WECC [21]:

$$\frac{dE'_q}{dt} = -\frac{1}{T_{p0}}[E'_q + (L_s - L_p)i_d] - \omega_0 \cdot s \cdot E'_d, \quad (1)$$

$$\frac{dE'_d}{dt} = -\frac{1}{T_{p0}}[E'_d - (L_s - L_p)i_q] + \omega_0 \cdot s \cdot E'_q, \quad (2)$$

$$\frac{dE''_q}{dt} = \left(\frac{1}{T_{pp0}} - \frac{1}{T_{p0}}\right)E'_q - \left(\frac{L_s-L_p}{T_{p0}} + \frac{L_p-L_{pp}}{T_{pp0}}\right)i_d + \omega_0 \cdot s \cdot (1 - T_{p0}) \cdot E'_d - \frac{1}{T_{pp0}}E''_q - \omega_0 \cdot s \cdot E''_d, \quad (3)$$

$$\frac{dE''_d}{dt} = \left(\frac{1}{T_{pp0}} - \frac{1}{T_{p0}}\right)E'_d + \left(\frac{L_s-L_p}{T_{p0}} + \frac{L_p-L_{pp}}{T_{pp0}}\right)i_q - \omega_0 \cdot s \cdot (1 - T_{p0}) \cdot E'_q - \frac{1}{T_{pp0}}E''_d + \omega_0 \cdot s \cdot E''_q, \quad (4)$$

$$\frac{ds}{dt} = -\frac{E''_d i_d + E''_q i_q - T_{m0}\omega_0^{Etrq}}{2H}, \quad (5)$$

$$i_d = \frac{R_s(V_d - E''_d) + L_{pp}(V_q - E''_q)}{R_s^2 + L_{pp}^2}, \quad (6)$$

$$i_q = \frac{R_s(V_q - E''_q) - L_{pp}(V_d - E''_d)}{R_s^2 + L_{pp}^2}, \quad (7)$$

$$p_{3\emptyset} = \frac{[R_s(V_d^2 + V_q^2 - V_d E''_d - V_q E''_q) - L_{pp}(V_d E''_q - V_q E''_d)]}{R_s^2 + L_{pp}^2}, \quad (8)$$

$$q_{3\emptyset} = \frac{L_{pp}(V_d^2 + V_q^2 - V_d E''_d - V_q E''_q) - R_s(V_d E''_q - V_q E''_d)}{R_s^2 + L_{pp}^2}. \quad (9)$$

where $E'_q$ and $E'_d$ are the transient voltages for IM on q-axis and d-axis. $E''_q$ and $E''_d$ represent the sub-transient voltages for IM on q-axis and d-axis. $T_{p0}$ and $T_{pp0}$ refer to the transient open-circuit time constant. $L_s$, $L_p$, and $L_{pp}$ indicate the synchronous reactance, transient reactance, and sub-transient reactance. Stator resistance is denoted by $R_s$.

Each of the three-phase induction motors represents a specific type of dynamic load. According to [26], **Ma** indicates the aggregation of the three-phase motor's driving constant torque loads, such as commercial/industrial air conditioner; **Mb** represents the aggregation of the three-phase motor's driving torque speed-squared loads with high inertia, such as fan motors used in residential and commercial buildings; **Mc** refers to the aggregation of three-phase motor's driving torque speed-squared loads with low inertia, such as direct-connected pump motors used in commercial buildings. Several technical reports [26], [27] have published their parameter settings for WECC CLM. However, those suggested parameters cannot accurately adapt and approximate every real-world case. Therefore, we design a variation range for each parameter based on [26] and assume the true values of these load parameters should fall into this range. Table I presents part of the designed parameter variation range for **Ma**, **Mb**, and **Mc**. In the first stage of our load modeling framework, which is the load composition identification, the load parameters of each load component are unknown and randomly selected from the designed range.

TABLE I
PARAMETER VARIATION RANGE FOR INDUCTION MOTOR

| Parameter | Ma | Mb | Mc |
|---|---|---|---|
| $R_s$ | [0.03, 0.05] | [0.03, 0.05] | [0.03, 0.05] |
| $L_s$ | [1.50, 2.00] | [1.50, 2.00] | [1.50, 2.00] |
| $L_p$ | [0.10, 0.15] | [0.17, 0.22] | [0.17, 0.22] |
| $L_{pp}$ | [0.10, 0.20] | [0.12, 0.15] | [0.12, 0.15] |
| $T_{p0}$ | [0.09, 0.10] | [0.18, 0.22] | [0.18, 0.22] |
| $T_{pp0}$ | [1e-3, 2e-3] | [2e-3, 3e-3] | [2e-3, 3e-3] |
| $H$ | [0.10, 0.20] | [0.25, 1.00] | [0.10, 0.20] |

### C. Single-phase Induction Motor

The single-phase IM **Md** is developed based on extensive laboratory testing by WECC [21], which can model both the protective devices and the compressors. The motor's P and Q consumptions are modeled with exponential characteristics, which are divided into three states as functions of bus voltage. State 1 applies when the bus voltage is higher than the motor compressor breakdown voltage (p.u.): $V > V_{brk}$, as shown in (10) state 2 applies when the bus voltage is in between the motor compressor breakdown voltage and motor compressor stall voltage: $V_{stall} \leq V \leq V_{brk}$, which is shown in (11); and state 3 applies when the bus voltage is lower than the motor compressor stall voltage: $V < V_{stall}$, as shown in (12):

$$state\ 1: \begin{cases} p_{1\emptyset} = p_{0,zip} \\ q_{1\emptyset} = q_{0,1\emptyset} + 6 \cdot (V - V_{brk})^2, \end{cases} \quad (10)$$

$$state\ 2: \begin{cases} p_{1\emptyset} = p_{0,1\emptyset} + 12 \cdot (V_{brk} - V)^{3.2} \\ q_{1\emptyset} = q_{0,1\emptyset} + 11 \cdot (V_{brk} - V)^{2.5}, \end{cases} \quad (11)$$

$$state\ 3: \begin{cases} p_{1\emptyset} = \frac{V^2}{R_{stall}} \\ q_{1\emptyset} = -\frac{V^2}{X_{stall}} \end{cases}. \quad (12)$$

where $p_{0,1\emptyset}$ and $q_{0,1\emptyset}$ are initial active and reactive power consumed by the single-phase motor. $R_{stall}$ and $X_{stall}$ are the compressor stalling resistance and reactance, respectively. The compressor motors are classified into two categories depending on if they can restart or not after stalling. The active power $p_{1\emptyset}$ and reactive power $q_{1\emptyset}$ consumed by all the compressor motors before and after stalling are shown in (13) and (14). A denotes the compressor motors that can be restarted, and B marks those that cannot be restarted. In (13), $F_{rst}$ refers to the ratio between motor loads that can restart and the total motor loads. In (14), $V_{rst}$ refers to the restarting voltage threshold for the stalled motors. $f(V > V_{rst})$ is the function of the P, Q recovery rate of the compressor motors that can be restarted.

$$before\ stalling: \begin{cases} p_A = p_{1\emptyset} * F_{rst} \\ q_A = q_{1\emptyset} * F_{rst} \end{cases}, \quad (13)$$

$$after\ stalling: \begin{cases} p_{1\emptyset} = p_A \cdot f(V > V_{rst}) + p_{B,stall} \\ q_{1\emptyset} = q_A \cdot f(V > V_{rst}) + q_{B,stall} \end{cases}. \quad (14)$$

Other than the voltage stalling feature introduced here, WECC CLM also incorporates a thermal relay feature into the single-phase motor, and the detailed information can be found in [21]. **Md**'s compressor dynamic model is the same as the three-phase IM as **Ma**, **Mb**, and **Mc**. We design the parameter selection range for **Md** according to [26]. The values of some critical parameters such as $V_{stall}$, $V_{rst}$, $V_{brk}$, and $F_{rst}$ are selected from the ranges shown in Table II.

TABLE II
PARAMETER VARIATION RANGE FOR SINGLE-PHASE IM

| Parameter | $V_{brk}$ | $V_{rst}$ | $V_{stall}$ | $F_{rst}$ |
|---|---|---|---|---|
| | [0.85, 0.90] | [0.92, 0.96] | [0.55, 0.65] | [0.15, 0.30] |

### D. Static Load Model: ZIP

The standard ZIP model is used in WECC CLM to represent the static load. The corresponding active and reactive power are written in (15)-(17):

$$p_{zip} = p_{0,zip} \cdot \left( p_{1c} \cdot \left(\frac{V}{V_o}\right)^2 + p_{2c} \cdot \frac{V}{V_o} + p_{3c} \right), \quad (15)$$

$$q_{zip} = q_{0,zip} \cdot \left( q_{1c} \cdot \left(\frac{V}{V_o}\right)^2 + q_{2c} \cdot \frac{V}{V_o} + q_{3c} \right), \quad (16)$$

$$\begin{cases} p_{1c} + p_{2c} + p_{3c} = 1, (0 \leq p_{1c}, p_{2c}, p_{3c} \leq 1) \\ q_{1c} + q_{2c} + q_{3c} = 1, (0 \leq q_{1c}, q_{2c}, q_{3c} \leq 1) \end{cases}. \quad (17)$$

where, $p_{0,zip}$ and $q_{0,zip}$ are the initial active and reactive power consumed by the ZIP load. $p_{1c}, p_{2c}$, and $p_{3c}$ are the coefficients for the active power of constant impedance, constant current, and constant power load. $q_{1c}, q_{2c}$, and $q_{3c}$ are the coefficients for reactive power of constant impedance, constant current, and constant power load. To model the diversity of ZIP load, the $p_{1c,2c,3c}$ and $q_{1c,2c,3c}$ are set to be random within the boundary shown in (17).

### E. Electronic Load

The electronic load model in the WECC CLM aims to simulate the linear load tripping phenomenon of electronics. It is modeled as a conditional linear function of the bus voltage $V$, as shown from the (18)-(19). $V_{d1}$ represents the voltage threshold at which the electronic load starts to trip, $V_{d2}$ indicates the voltage threshold at which all the electronic load trips, $V_{min}$ tracks the minimum bus voltage during the transient, $frcel$ indicates the fraction of electronic load that can be restarted after a fault is cleared. In (20), $pf_{elc}$ denotes the power factor of electronic load (default as 1), and $p_{0,elc}$ refers to the initial power of electronic load. The parameter variation ranges for electronic load are the shown in Table III.

$$fvl = \begin{cases} 1 \\ \frac{V - V_{d2}}{V_{d1} - V_{d2}} \\ \frac{V_{min} - V_{d2} + frcel \cdot (V - V_{min})}{V_{d1} - V_{d2}} \\ 0 \end{cases}, \quad (18)$$

$$p_{elc} = fvl \cdot p_{0,elc}, \quad (19)$$

$$q_{elc} = \tan(\cos^{-1}(pf_{elc})) * p_{elc}. \quad (20)$$

TABLE III
PARAMETER VARIATION RANGE FOR ELECTRONIC LOAD

| Parameter | $V_{d1}$ | $V_{d2}$ | $pf_{elc}$ |
|---|---|---|---|
| | [0.60, 0.70] | [0.50, 0.55] | 1 |

### F. Identify the Composition of the Composite Load

In a composite load model, different load composition can induce very similar dynamic responses [22], [23]. It has been observed in [22] that a different load composition of a big IM and a small IM could have very similar load dynamic responses. This multi-solution phenomenon on load composition is even more common in the WECC CLM due to the multiple IMs in place. Our proposed two-stage load modeling method can effectively find near-optimal load compositions in stage one; and then in stage two, the other load parameters can be efficiently identified. To demonstrate the importance of identifying the load composition before fitting other parameters, we conduct a fitting loss comparison.

In this comparison, we first create one set of reference P, Q dynamic curves, and then according to the reference curves, we fit one load composition using our proposed load modeling method. Then, we use the true load composition and generate a random load composition as two comparison groups. We gradually increase the number of sampled load models under these three load compositions from one to one hundred. The mean fitting losses of these three load compositions are plotted and compared in Fig. 2. When the sample number is small, the fitting loss of the fitted load composition is similar to the random load composition, and the fitting loss of the true load composition is much lower. However, as the sample number increases, the mean fitting loss of the fitted load composition quickly decreases and eventually merges with the true load composition, but the fitting loss of the random load composition stays high. This comparison justifies the effectiveness of conducting load composition identification before fitting other parameters. The customized fitting loss is introduced in Section III. B.

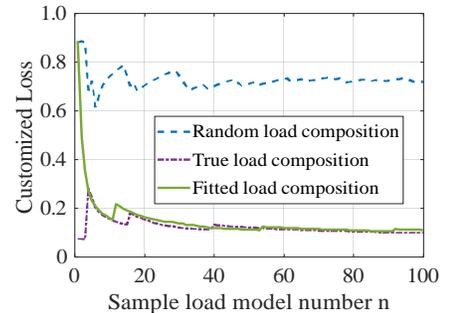

Fig. 2. Sensitivity analysis on sample number n against mean fitting loss

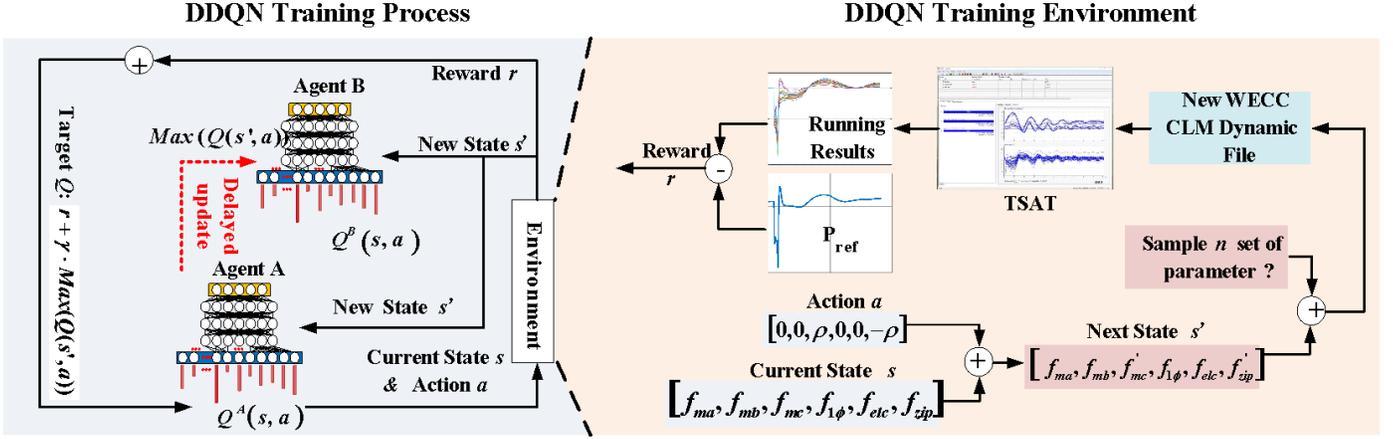

Fig. 3. The DDQN agent training process and training environment introduction

In stage one, the DDQN agent will find multiple high-quality load composition solutions due to their similar dynamic responses. We use pinball loss (or named as quantile loss) as a quantile-based metric to evaluate each load composition according to the produced dynamic response prediction intervals in a probabilistic manner [28]. Just like RMSE in the point forecasting, pinball loss, as shown in (21), calculates a value to indicate the accuracy of the generated quantile with the reference values [29]. The lower the loss is, the better the quantile is produced. In (21), $\hat{x}_o$ is the value at quantile $o$ of a group of data, $x$ indicates the value that need to be evaluated, and $\tau$ refers to the penalize factor at the corresponding quantile level. By calculating the mean pinball loss of both the $P_{ref}$ and $Q_{ref}$ within a quantile band $[\underline{\tau}, \overline{\tau}]$ generated by different load compositions, we can rank each load composition in representing the true load dynamics as shown in (22). Where, $P_o^i/Q_o^i$ are the value of $i^{th}$ snapshot of $P/Q$ dynamic response at quantile $o$ under load composition $S$, $P_{ref}^i/Q_{ref}^i$ show the $i^{th}$ snapshot of $P/Q$ reference dynamic response, $\overline{\tau}$ is the penalize factor for upper bound quantile and $\underline{\tau}$ is the penalize factor for quantile lower bound, $N$ refers to the number of snapshots in the dynamic curves:

$$L_\tau(\hat{x}_o, x) = \max[(\hat{x}_o - x) \cdot \tau, (\hat{x}_o - x) \cdot (\tau - 1)] \quad (21)$$

$$L^{pinball} = \frac{\sum_{i=1}^N [L_{\overline{\tau}}(P_o^i, P_{ref}^i) + L_{\underline{\tau}}(P_{1-o}^i, P_{ref}^i) + L_{\overline{\tau}}(Q_o^i, Q_{ref}^i) + L_{\underline{\tau}}(Q_{1-o}^i, Q_{ref}^i)]}{N} \quad (22)$$

To get the quantile value under each load composition, transient responses of the WECC CLM models with the same load composition but different remaining parameters are generated. An example is given in Fig. 4 to show value intervals of $P$ response under a certain load composition using 500 random cases. These 500 cases are randomly generated by uniformly sampling within the defined parameter boundaries. According to our tests, 500 cases is sufficient to form a representative distribution on the value intervals of each snapshot. The value distributions of two snapshots are presented in Fig. 4.

### G. Monte Carlo-based Parameter Selection

In the last step, the probability of each possible load composition is calculated using (22). Then, from the massive random cases that are used to generate the distributing band as shown in Fig. 4, the set of parameters that best approximates the reference dynamics $P_{ref}$ and $Q_{ref}$ is selected as the load modeling result. The fitting accuracy is measured using Root Mean Squared Error (RMSE).

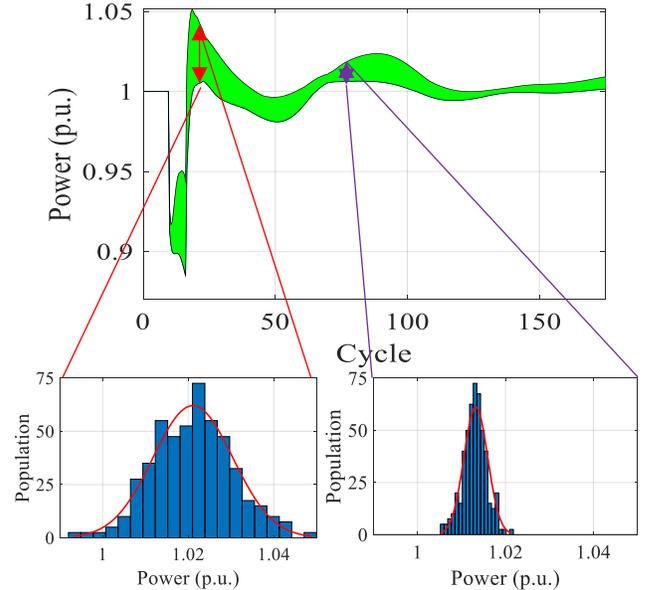

Fig. 4. Dynamic response value intervals under a load composition at different snapshots

### III. DDQN-BASED LOAD COMPOSITION IDENTIFICATION

#### A. DDQN Agent Training Setup

In recent years, AI embraces a giant development. Lots of AI techniques are studied and implemented in power system to address the complex control problems [30] - [32], which were hard to be solved using conventional techniques, and load modeling for WECC CLM is one of them. In this paper, we leverage DDQN technique to solve this problem. In DDQN two neural network agents are trained to interact with the environment. Agent A is the prediction network that performs the actions to the environment and updates at each training step, and agent B is the target network which provides a target Q value for agent A's updating while agent B is updated at every $C$ steps ($C \gg 1$). Compared to the regular DQN algorithm,

DDQN has better training stability as it avoids the positive bias propagation caused by the max function in a Bellman equation [33]. At each state, the environment responds to the taken action. This response is interpreted as reward or penalty. Both agent A and agent B learn the action-reward function $Q(s,a)$ by iteratively updating the $Q$ value following (23), which is fundamentally a Bellman equation. In (23), the $Q^A(s,a)$ and $Q^B(s,a)$ denote the $Q$ functions learned by agent A and agent B; $s$ is the current state; $a$ refers to the current action taken by the agent. $\delta$ represents the learning rate, which determines to which extent the newly acquired information overrides the old information. $\gamma$ indicates the discount factor, which essentially determines how much the reinforcement learning agent weights rewards in the long-term future relative to those in the immediate future. $r$ is the immediate reward/penalty by taking action $a$ at state $s$; $s'$ is the new state transient from $s$ after action $a$ is taken.

$$Q^A(s,a) = (1-\alpha)Q^A(s,a) + \delta \cdot (r + \gamma \cdot \max Q^B(s',a)) \quad (23)$$

Function $Q^A(s,a)$ updates at every step following (23), but function $Q^B(s,a)$ updates every $C$ steps ($C \gg 1$). In such a way, the temporal difference (TD) error is created, which serves as the optimization target for the agent, as shown in (24).

$$\min(\mathcal{L}) = \|Q^A(s,a) - r - \gamma \cdot \max Q^B(s',a)\| \quad (24)$$

In this application, the state is defined as the load composition fraction of each load component: $s = [f_{ma}, f_{mb}, f_{mc}, f_{1\phi}, f_{elc}, f_{zip}]$. The summation of $s$ is always one to represent the full load. The actions to be taken by the agents are the pair-wise load fraction modification: $a = [\cdots, \rho, \cdots, -\rho, \cdots]$. $\rho$ is the fraction modification value, which is designed as 0.01 in the case study. Each $a_t$ only has two non-zero elements, which are $\rho$ and $-\rho$. In this case, the summation of $s$ is guaranteed to remain one at each step. For WECC CLM in the study, there are six load components. Considering the fraction has plus/minus two directions to update, the total number of two-combinations from six elements is $A_6^2 = 6 \times 5 = 30$. The training environment is the IEEE 39-bus system built in the Transient Security Assessment Tool (TSAT) in DSATools™. Fig. 3 shows the DDQN training process and the training environment. Observed from the training environment, when a new state $s'$ is reached, $n$ sets of parameters θ will be sampled, which are then combined with $s'$ to form $n$ dynamic files. The $n$ dynamic files are run in the TSAT in order to calculate the reward. In our work, $n$ is selected as 20 to efficiently identify the good load composition candidates through the sensitivity analysis shown in Fig. 2.

---

**Algorithm I**: DDQN Training for WECC CLM

**Input**: Reference dynamic responses $P_{ref}$ and $Q_{ref}$.
**Output**: Load composition and load parameters
Initialize $\lambda, \gamma, \varepsilon, \eta$, NN. A, NN. B and memory buffer $M$
**For** i **in** range (number of episode)
    $s$←reset.enviroment(); $\varepsilon$← $\varepsilon \cdot \eta$; r_sum←0; tik←0; NN. B←NN. A;
    **While** $tik \leq 80$:
        **If** rand(1) < $\varepsilon$:
            $a$← $a$(randi(|30|))
        **Else**:
            $a$← $a$(argmax(NN. A. predict($s$)))
        **End**
        $s', r$←execute.TSAT($s, a$)
        **If** r>λ
            Terminate Episode i.
        **Else**
            *Step1*: $Q^B(s,a) = $ NN. B. predict($s, a$)
            *Step2*: $Q^A(s) = $ NN. A. predict($s$)
            *Step3*: $Q^A(s)(\text{index}(a \text{ in } a)) = Q^B(s,a) + r$
            Sample a batch of transitions $D$ from $M$
            Repeat the *Step* 1 to *Step* 3 for each sample in $D$.
            NN. A. fit([$s, s_D$],[$Q^A(s), Q_D$])
            $M$←[$s, a, s', r$]
            $s \leftarrow s'$
            r_sum= r_sum+r
        **End**
    r_list.append(r_sum)
**End**

---

The pseudo-code for the DDQN agent training is shown in Algorithm I. In the training process the epsilon-greedy searching policy and the memory replay buffer are applied, and the detailed introduction to them can be found from [34], [35], which will not be discussed in this paper. In our application, the memory buffer size is designed as 2,000.

### B. Customized Reward Function

The reward in our application is a negative value that represents the transient P and Q curve fitting losses. The training goal is to maximize the reward in (25) or equivalently minimize the fitting losses. A higher reward means a higher fitting accuracy. At each new state, the dynamic responses are compared with the reference responses to get a reward $r$, which will be further interpreted into a Q value to update the agent A and agent B. However, the classic RMSE loss function cannot properly differentiate the desirable load compositions from the undesirable ones. This phenomenon is further explained later. Therefore, a customized loss function is developed to better capture the dynamic features of the transient curves as shown in (25) and (26):

$$r = -\alpha \cdot r_{RMSE} - \beta \cdot r_{trend} - r_{step}. \quad (25)$$

$$r_{trend} = \frac{\sum_{i=1}^{n} |idx_{min}^i - idx_{min}^{ref}| + |idx_{max}^i - idx_{max}^{ref}|}{K} \quad (26)$$

where $r_{RMSE}$ denotes the RMSE between $\boldsymbol{P}_{test}, \boldsymbol{Q}_{test}$ and $\boldsymbol{P}_{ref}, \boldsymbol{Q}_{ref}$. In equation (25), the regularization term $r_{trend}$ represents the time index mismatch of peak and valley values between $\boldsymbol{P}_{test}, \boldsymbol{Q}_{test}$ and $\boldsymbol{P}_{ref}, \boldsymbol{Q}_{ref}$. $\alpha$ and $\beta$ are the weights of term $r_{RMSE}$ and $r_{trend}$. In equation (26), $K$ is a constant that scales down the index mismatch between $\boldsymbol{P}_{test}, \boldsymbol{Q}_{test}$ curves and the $\boldsymbol{P}_{ref}, \boldsymbol{Q}_{ref}$, $idx_{min/max}^i$ refers to the index of the minimum/maximum value in the $i^{\text{th}}$ $\boldsymbol{P}_{test}, \boldsymbol{Q}_{test}$, and $idx_{min/max}^{ref}$ indicates the index of the minimum/maximum value of $\boldsymbol{P}_{ref}, \boldsymbol{Q}_{ref}$. The values of $\alpha, \beta$, and $K$ are tuned so that the value of $-\alpha \cdot r_{RMSE} - \beta \cdot r_{trend} - r_{step}$ is normalized into the range of [-1, 0]. This term explicitly differentiates the desirable fitting results from others and enforces the similar peak and valley timestamps as $\boldsymbol{P}_{ref}, \boldsymbol{Q}_{ref}$. Another regularization term $r_{step}$ is a constant penalty for each step of searching, which facilitates the agent's training speed. Such loss function is fundamentally a similarity-based measure, and this type of metrics is commonly used in load modeling techniques [36]. By using this customized loss function, a generic fitting accuracy threshold λ can then be set as the episode termination condition. We show some example plots in Fig. 5 to better explain the effects of this customized loss function.

In Fig. 5, the $\boldsymbol{P}_{ref}$ is a $P$ dynamic response from a WECC CLM, located at bus 20 of the IEEE 39-bus system, and a three-phase fault is deployed at bus 6. The plots are normalized based on the power flow solution at steady state; Reference Group* shows multiple $\boldsymbol{P}$ dynamic responses from multiple WECC CLMs that have the same load composition as the $\boldsymbol{P}_{ref}$, but with different load parameters. The other four plots are called the comparison groups, where the transient $\boldsymbol{P}$ curves in each plot are generated by the WECC CLMs with a different load composition.

The RMSE and customized loss between the $\boldsymbol{P}_{ref}$ and these five groups are summarized in Table IV. It shows that the RMSEs of the five groups are very close. The boundary between the desirable composition and the undesirable compositions is not clear. Motivated by the aim of load modeling to replicate key features from dynamic responses, we designed our customized loss function to better differentiate the good composition from the less good ones. In this case, it is difficult to derive a generic threshold λ for the DDQN algorithm that is applicable to all cases. On the contrary, by using the customized loss function, the fitting loss discrepancy between the Reference Group* and other groups are significantly enlarged as shown in Table IV. As a result, a generic and fixed λ can be defined to serve as the termination condition for each episode of training.

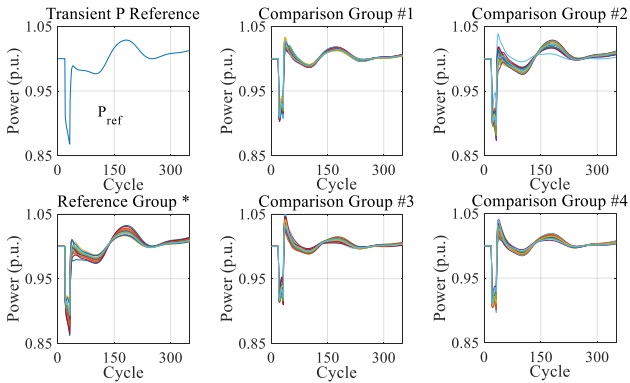

Fig. 5. Loss comparison between reference group and comparison groups

TABLE IV
PARAMETER VARIATION RANGE FOR ELECTRONIC LOAD

|  | Reference Group* | Group 1 | Group 2 | Group 3 | Group 4 |
|---|---|---|---|---|---|
| **RMSE** | 0.0136 | 0.0353 | 0.0205 | 0.0413 | 0.0338 |
| **Customized Loss** | -0.0078 | -0.9596 | -0.1702 | -0.9996 | -0.7206 |

## IV. CASE STUDIES

### A. Test Environment

The transient stability test cases shown in this section are conducted in IEEE 39-bus system. In each case study, the base contingency is chosen as a three-phase fault occurred at bus 6, and the load model to be identified is located at bus 20. All the cases are performed using the Transient Security Assessment Tool (TSAT) in DSATools™ developed by Powertech Labs Inc.

### B. Case I: Algorithm Test on CLM with ZIP + IM

In Case I, the performance of the proposed algorithm is tested on the conventional ZIP + IM composite load model (CLM). For the DDQN agent, the state vector $s$ indicates the composition of the two load types $\boldsymbol{s} = [s_{zip}, s_{IM}]^T$, ($s_{zip} + s_{IM} = 1$). Since there are only two load components to be identified, the action space only contains two actions, which are $\boldsymbol{a} = [a_1, a_2]^T = \begin{bmatrix} 0.01 & -0.01 \\ -0.01 & 0.01 \end{bmatrix}$.

The reference load composition is $\boldsymbol{s}_{ref} = [0.2937, 0.7063]^T$. The DDQN agent starts to search for possible solutions from a randomly generated load composition [0.4935, 0.5065]. The agent training process is shown in Fig. 6. The training reward converges after around 2,000 episodes. The top 3 most possible solutions selected by the trained DDQN agent are listed in Table V and their corresponding $P$ dynamic responses are plotted in Fig. 7. All the three solutions found by the agent have very similar dynamic responses with the actual load model.

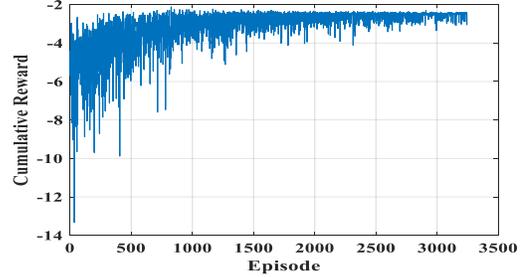

Fig. 6. DDQN learning process for ZIP+IM load model

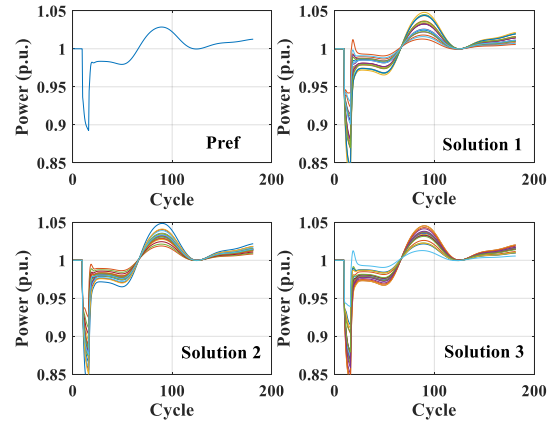

Fig. 7. Reference P curve and the top three possible solutions

The prediction intervals of the three solutions are calculated using pinball loss with quantile interval as 70 % coverage at [15%, 85%], 80 % coverage at [10%, 90%], and 90 % coverage at [5%, 95%]. The results are also listed in Table V. Among the three solutions, solution 2 has consistently achieved the lowest pinball loss for all the three quantile intervals. Therefore, it is selected as the load composition identification solution: $S = [0.2935, 0.7065]^T$.

TABLE V
CANDIDATE LOAD COMPOSITION

| | | True | Solution 1 | Solution 2 | Solution 3 |
|---|---|---|---|---|---|
| ZIP | | 0.2937 | 0.2835 | 0.2935 | 0.3035 |
| IM | | 0.7063 | 0.7165 | 0.7065 | 0.6965 |
| Pinball Loss | [15%, 85%] | | 0.0306 | **0.0286** | 0.0338 |
| | [10%, 90%] | | 0.0295 | **0.0273** | 0.0328 |
| | [5%, 95%] | | 0.0295 | **0.0269** | 0.0331 |

Based on the solution, 500 Monte Carlo samplings are conducted on the load parameters. The one set of parameters, yielding the lowest dynamic response reconstruction error, is selected as the identified load parameters. The reference load parameters and the identified load parameters are shown in Table VI. Except $P_{1C}$ and $P_{2C}$, all the other parameters are well fitted. The P and Q transient dynamic response comparisons between the reference model and the identified model are shown in Fig. 8. The active power P fitting RMSE is 0.0692% and the reactive power Q fitting RMSE is 0.68%.

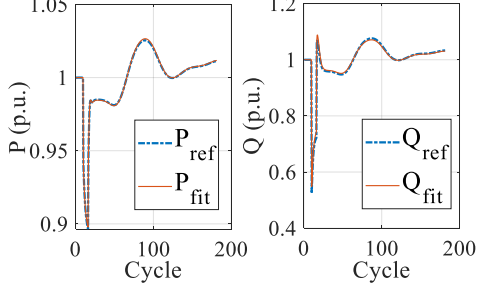

Fig. 8. Dynamic responses comparison between the reference load and the fitted load.

TABLE VI
CANDIDATE LOAD COMPOSITION

| | $R_s$ | $L_s$ | $L_p$ | $L_{pp}$ | $T_{p0}$ | $T_{pp0}$ |
|---|---|---|---|---|---|---|
| Ref | 0.0314 | 1.9013 | 0.1228 | 0.1040 | 0.0950 | 0.0021 |
| Fit | 0.0327 | 1.8558 | 0.1328 | 0.1032 | 0.0938 | 0.0021 |
| | H | Etrq | $P_{1C}$ | $P_{2C}$ | $Q_{1C}$ | $Q_{2C}$ |
| Ref | 0.1000 | 0 | 0.0316 | 0.6947 | -0.4769 | 1.4769 |
| Fit | 0.1030 | 0 | 0.0274 | 0.2287 | -0.4477 | 1.4477 |

### C. Case II: Algorithm Test on WECC CLM

In this case, the proposed DDQN-based load composition identification strategy is applied to the WECC CLM. Compared with Case I, the number of load component in the WECC CLM increases from two to six. Therefore, the state vector size turns into 6×1. The number of actions can be taken by the agent also increases to $A_6^2=30$. The action step size is 0.01, which means the load composition changes 1% at each step. This case study aims to demonstrate that the proposed method is scalable to larger load models.

The reference load composition is $s_{WECC} = [0.3637, 0.1430, 0.0914, 0.1526, 0.1088, 0.1405]^T$. The training starting state is defined as $[1/6, 1/6, 1/6, 1/6, 1/6, 1/6]^T$. The training reward converges after 900 episodes as shown in Fig. 9. The scatter plot in Fig. 9 shows the summation of P and Q fitting RMSE at each training episode, which converges at the same pace with the cumulative reward. The fraction evolutions of the six load components are also plotted in Fig. 10. It shows that at the beginning of the training process, the DDQN agent actively explores the state space and leads to a large search variance for each load component. As the training proceeds, the searching range of the DDQN agent becomes more concentrated and eventually converges to finish the exploitation. This figure verifies the strong exploration and exploitation capabilities of the proposed approach. The top three most possible solutions given by the agent are listed in Table VII and their corresponding P dynamic responses are plotted in Fig. 11.

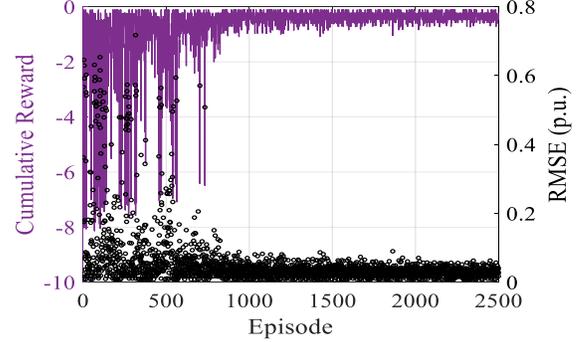

Fig. 9. DDQN learning process for WECC CLM

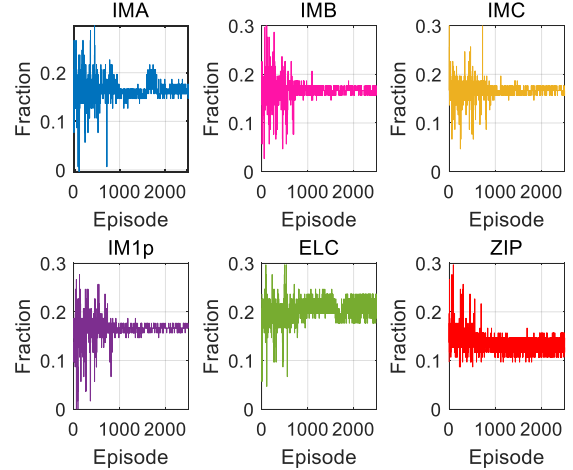

Fig. 10. Evolution of load fractions for the six load components over the training

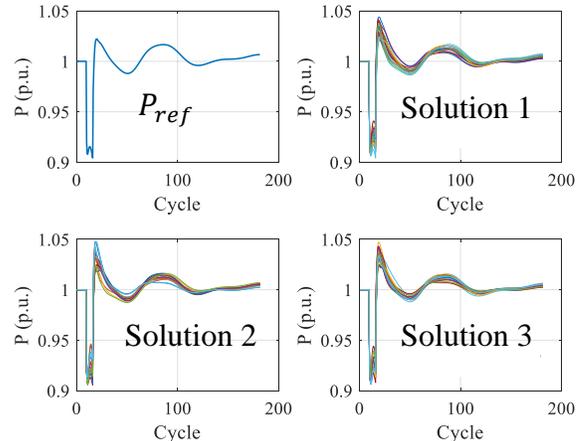

Fig. 11. Reference P curve and the top three possible solutions

Unlike the conventional CLM with only one IM, the WECC

CLM has three IMs and one single-phase IM; therefore, the transient dynamics between each load component has more mutual interference. For each transient event, there exist multiple load composition solutions with very similar transient dynamics [23]. As shown in Table VII, the top three most possible solutions are listed. For those three solutions, the load distribution among dynamic loads and static loads are close to the reference load model. During the training process, the DDQN agent gradually learns to choose solutions with lower fitting quantile loss; in other words, a more stable solution emerges so that each episode is terminated with fewer exploration steps. According to the lowest pinball loss at different percentile intervals, solution 1 is chosen as the load composition solution. Based on this result, 500 Monte-Carlo samplings are conducted to select a set of parameters that best match with the reference P and Q. The best fitting result is shown in Fig. 12. Due to space limitation, the parameters of the reference load and identified load are not presented.

Noted, the initial state is selected assuming no prior information about the load composition. When there is previous load statistics, a better initial state can be derived.

TABLE VII
CANDIDATE LOAD COMPOSITION

|  |  | True | Solution 1 | Solution 2 | Solution 3 |
|---|---|---|---|---|---|
|  | IM_A | 0.3637 | 0.1667 | 0.1667 | 0.1767 |
|  | IM_B | 0.1430 | 0.1667 | 0.1567 | 0.1567 |
|  | IM_C | 0.0914 | 0.1667 | 0.1667 | 0.1667 |
|  | IM_1p | 0.1526 | 0.1667 | 0.1767 | 0.1567 |
|  | ELC | 0.1088 | 0.2067 | 0.2167 | 0.2267 |
|  | ZIP | 0.1405 | 0.1267 | 0.1167 | 0.1167 |
|  | Dynamic | 0.7507 | 0.6667 | 0.6667 | 0.6566 |
|  | Static | 0.2493 | 0.3333 | 0.3333 | 0.3434 |
| Pinball Loss | [0.15,0.85] |  | **0.0143** | 0.0153 | 0.0222 |
|  | [0.10,0.90] |  | **0.0136** | 0.0147 | 0.0185 |
|  | [0.05,0.95] |  | **0.0131** | 0.0140 | 0.0162 |

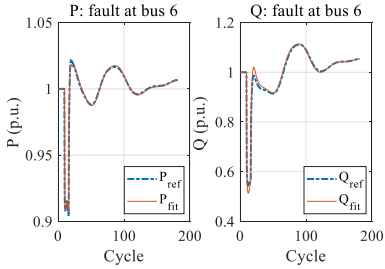

Fig. 12. Dynamic responses comparison between the reference load and the fitted load.

### D. Case III: Model Robustness Tests

One of the most important reasons for load modeling is to have a consistent load representation that can closely reflect the real transient dynamics under different contingencies. For that purpose, another three groups of robustness tests are simulated. In the first group, the fault location is changed from bus 1 all the way up to bus 39. In the second group, the fault type is modified from three-phase fault to single-phase-to-ground fault and double-phase-to-ground fault. In the third group, the fault duration is changed from the original 6 cycles (100 ms) to 8 cycles (133.33 ms) and 10 cycles (166.67 ms).

The results of the first group of tests show that when the fault occurs at other buses, the $P$, $Q$ transient curves of the identified load model still fit the true transient curves very well. Fig. 13 shows the $P$, $Q$ transient examples for faults that occur at bus 14 and bus 29, respectively. In this group of tests, the active power $P$'s fitting RMSE has a mean value of 0.0995% (0.0255% $\leq$ $RMSE_P \leq$ 0.2124%). For reactive power $Q$, the mean fitting RMSE is 0.7852% (0.2374% $\leq RMSE_Q \leq$ 1.5939%). The high dynamic fitting accuracy achieved by the fitted load model demonstrates the proposed load modeling method's robustness towards faults that occur at different locations.

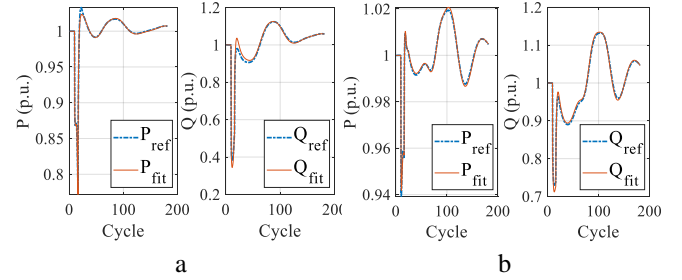

Fig. 13. P and Q fitting comparisons when fault occurs at (a) bus 14. (b) bus 29

The results of the second group of tests show that the identified load model can capture the transient behaviors of the reference load model under different fault types. Fig. 14 shows the $P$, $Q$ fitting curves of our identified load model when single-phase-to-ground fault and double-phase-to-ground fault occur at bus 6. The same test at other buses are also conducted. In summary, the mean $P$ fitting RMSE is 0.0714% (0.0236% $\leq RMSE_P \leq$ 0.1447%); the mean $Q$ fitting RMSE is 0.7216% (0.2111% $\leq RMSE_Q \leq$ 1.3372%). This test demonstrates the robustness of the proposed load modeling method towards different fault types. The case study also proves the scalability of the method to larger load models.

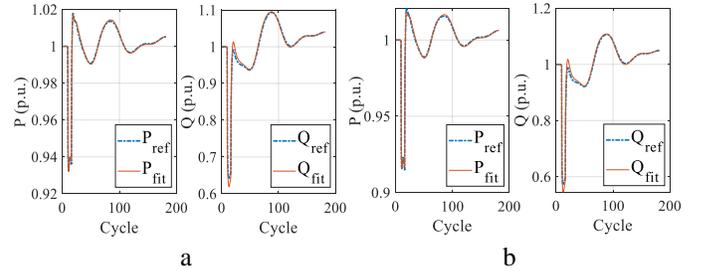

Fig. 14. P and Q fitting comparisons for (a). single-phase-to-ground fault. (b) double-phase-to-ground fault.

The results of the third group of tests show that the identified load model can reproduce the responses of the reference load model with different fault durations as well. During the identification phase, the reference event fault is cleared with 6 cycles. Then, longer-duration faults are used to test the robustness of the identified model. Fig. 15 (a) and (b) show the P, Q fitting curves when the fault occurs at bus 6 for 8 and 10 cycles. The same test on other buses is also conducted. In summary, when the fault lasts 8 cycles, the mean P fitting RMSE is 0.1008% (0.07641% $\leq RMSE_P \leq$ 0.1974%); and the mean Q fitting RMSE is 0.8566% (0.5133% $\leq RMSE_Q \leq$ 1.7712%). When the fault lasts for 10 cycles, the mean P fitting RMSE is 0.1804% (0.1236% ≤ $RMSE_P$ ≤0.2113%); and the mean Q fitting RMSE is 1.2677% (0.7323% ≤ $RMSE_Q$ ≤ 1.8522%). The load profiles are well represented by the identified model, which demonstrates the robustness of the proposed method towards different fault durations.

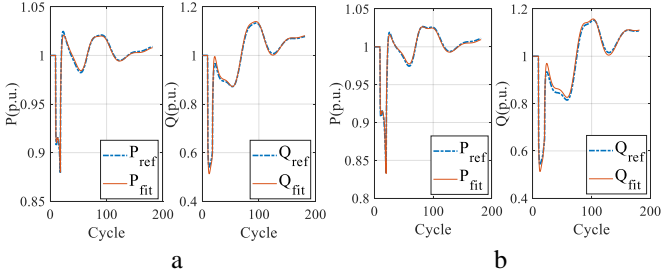

Fig. 15. P and Q fitting comparisons for (a). 8-cycle fault at bus 6. (b) 10-cycle fault at bus 6.

### E. High Penetration of Single-phase Induction Motor Load

WECC CLM is widely recognized for its capability of modeling the fault induced delayed voltage recovery (FIDVR) event, which is caused by the quickly changing real and reactive power load due to single-phase motor A/C stalling [37] . To simulate an FIDVR event at bus 20, a reference WECC CLM model is generated with a high single-phase induction motor load: $[0.1,0.1,0.1,0.52,0.1,0.08]^T$ . Then, a three-phase fault occurs at bus 32 at the 9$^{th}$ cycle and clears at the 21$^{st}$ cycle. Fig. 16 shows the bus voltage measured at bus 20 when the fault occurs. In this example, the voltage recovers at the 89$^{th}$ cycle which is 68 cycles delayed. Based on this FIDVR event, we test the performance of our proposed method for the case with high penetration of single-phase induction motors. Following the same initialization procedure as previous tests, we set the initial load fractions to be $[\frac{1}{6},\frac{1}{6},\frac{1}{6},\frac{1}{6},\frac{1}{6},\frac{1}{6}]^T$. As shown in Fig. 17(a), the training process converges after around 800 episodes, and the final identified load composition solution is $[0.1067, 0.0667, 0.1167, 0.3967, 0.1467, 0.1667]^T$. The dominance of the single-phase induction motor load is identified with the proposed DDQN method. The P, Q curve fitting results are shown in Fig. 17(b) with a fitting RMSE at 0.28% and 0.70%, respectively. This case study demonstrates the effectiveness of the performance of our proposed method under an FIDVR event.

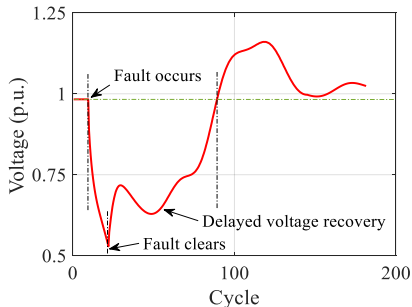

Fig. 16. Voltage profile for bus 20 during a FIDVR fault.

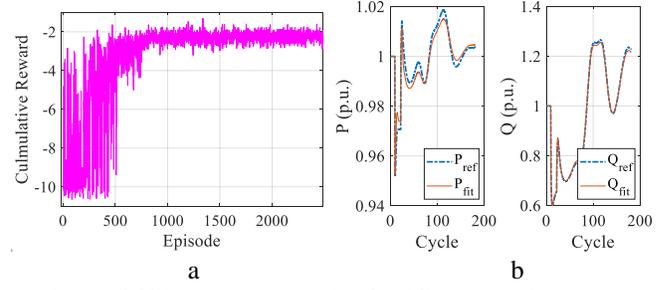

Fig. 17. (a). DDQN training process. (b). P and Q fitting results.

### F. Performance Comparison

To evaluate the performance of DDQN against other heuristic optimization algorithms, we apply particle swarm optimization (PSO) and genetic algorithm (GA) to optimize the load composition in stage one by using the same reward function. The reference WECC CLM is the same as the one in Case II. Unlike the proposed DDQN method with one single starting point, 30 initial PSO particles are randomly generated and the initial value for each load component is within the range of $\frac{1}{6} \pm \frac{3}{100}$. The result of PSO is shown in Fig. 18. Fig. 18a shows the load composition searching reward, and it stops increasing after 60 iterations. The converged reward is -0.0342 which is much worse than the DDQN fitting accuracy threshold λ (λ=-0.012), as higher values indicates better performances. The best load composition found by PSO is $[0.4536,0,0.0226,0.1694,0.1582,0.1964]^T$ . Based on this load composition, the Monte-Carlo simulation identified result is shown in Fig. 18b.

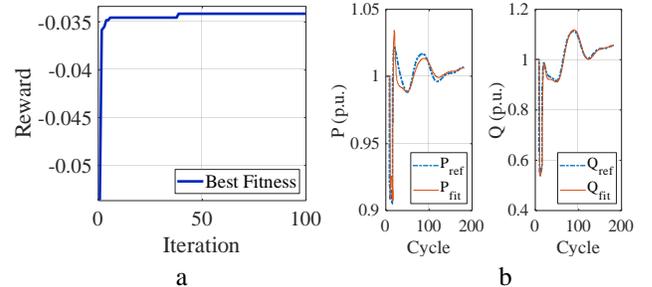

Fig. 18. (a) PSO-based load composition searching. (b) Dynamic responses comparison between the reference load and the PSO fitted load.

For the GA, the same 30 initial PSO particles are adopted as the first generation of GA parents. At each following generation, the top 30 offspring will be selected to reproduce. Fig. 19 shows the simulation results of GA. In Fig. 19a, the plot indicates that the best load composition searching reward reaches -0.0386. In general, the GA's performance is worse than PSO. The best load composition found by GA is $[0.0905, 0.1017, 0.1753, 0.2351, 0.2457, 0.1517]^T$ . Based on this load composition, the Monte-Carlo simulation finds the best fitting results as in Fig. 19b.

Table VIII summarizes the P, Q fitting accuracy using the load compositions found by PSO, GA, and DDQN. The proposed DDQN method outperforms PSO and GA by achieving the lowest RMSE. For PSO, its Q fitting accuracy is the same as DDQN. However, its P fitting accuracy is much worse than DDQN. GA has the worst P, Q fitting performances in this case. Since the second-stage parameter identification follows the same procedure for these three methods, this comparison also partially verifies our claims in Section II that identifying a proper load composition can greatly improve the dynamic response reconstruction efficiency.

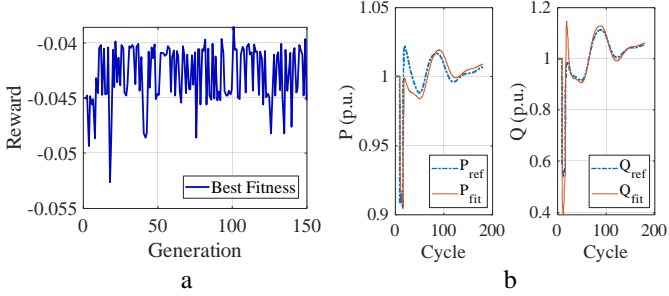

Fig. 19. (a) GA-based load composition searching. (b) Dynamic responses comparison between the reference load and the GA fitted load.

TABLE VIII
PERFORMANCE COMPARISON

|  | PSO | GA | DDQN |
|---|---|---|---|
| RMSE for P | 0.0050 | 0.0075 | **0.0012** |
| RMSE for Q | 0.0065 | 0.0358 | **0.0064** |

We conducted other two groups of comparison between PSO, GA, and DDQN, the results consistently show that DDQN's performance are better than PSO and GA, and PSO's performance is better than GA. This comparison also verifies our claims in Section II, that identifying a proper load composition can greatly improve the parameter fitting efficiency.

### G. Impact of Initial Point on the Algorithm Performance

The proposed load modeling method nonlinearly optimizes the load compositions. It is critical to evaluate the impacts of the initial point selection on the identification results. In this section, we design another WECC CLM with a reference load composition as $s_{WECC} = [0.1, 0.15, 0.1, 0.2, 0.1, 0.35]^T$. Then we conduct the two load modeling tests, Test-Rand and Test-Close, using two different initial points. The initial point for Test-Rand is the same as the Case II, which is $[1/6, 1/6, 1/6, 1/6, 1/6, 1/6]^T$. The initial point for Test-Close is designed to be very close to the reference load composition, which is $[0.08, 0.1, 0.13, 0.22, 0.07, 0.4]^T$. By comparing Test-Rand with Test-Close, we can evaluate the impacts of different initial points on the same case.

TABLE IX
PERFORMANCE COMPARISON

|  |  | True | Test-Rand | Test-Close |
|---|---|---|---|---|
| Stage one | IM_A | 0.1000 | 0.0967 | 0.0900 |
|  | IM_B | 0.1500 | 0.1667 | 0.1500 |
|  | IM_C | 0.1000 | 0.1667 | 0.0800 |
|  | IM_1p | 0.2000 | 0.1467 | 0.2100 |
|  | ELC | 0.1000 | 0.1667 | 0.0700 |
|  | ZIP | 0.3500 | 0.2567 | 0.4000 |
|  | Static Load | 0.4500 | 0.4234 | 0.4700 |
|  | Dynamic Load | 0.5500 | 0.5766 | 0.5300 |
|  | Pinball Loss |  | 0.0134 | 0.0125 |
| Stage two | P fitting RMSE |  | 0.0011 | 0.0013 |
|  | Q fitting RMSE |  | 0.0046 | 0.0019 |

Table IX shows the identified load compositions and P, Q fitting RMSE of Test-Rand and Test-Close. Fig. 20 shows the P, Q fitting curves for Test-Rand and Test-Close. Both Test-Rand and Test-Close achieve good load modeling results not matter for pinball loss or RMSE. Test-Rand is slightly better than Test-Close in P fitting RMSE, but slightly worse in Q fitting RMSE. Overall, two identified solutions are reasonably close to the true compositions. As load is constantly changing, how to obtain a "close" initial point is also non-trivial. It is good to have a close initial point to start with, but it is not required. The proposed method can effectively identify the load model and its parameters through proposed DDQN method.

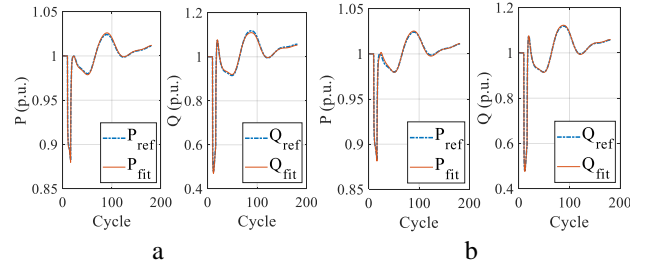

Fig. 20. Fitted P, Q curves for (a). Test-Rand. (b). Test-Close.

### H. Fit ZIP+IM and CLOD Using WECC CLM

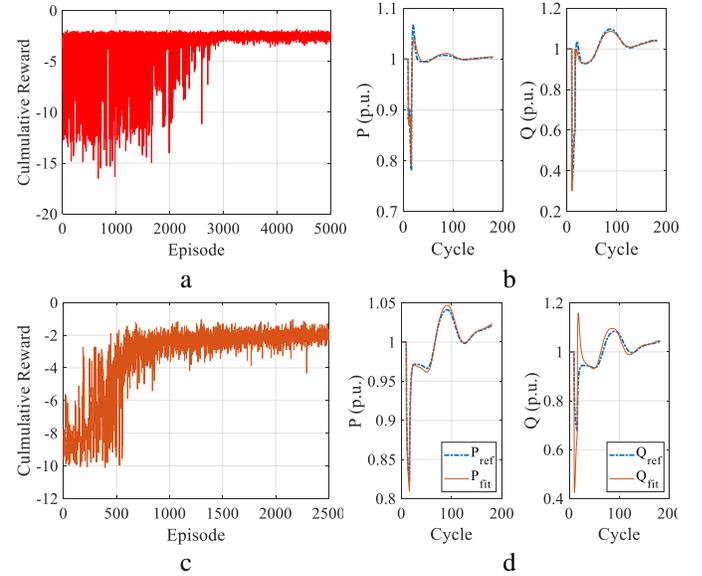

Fig. 21. (a) DDQN learning process for fitting ZIP+IM's dynamics using WECC CLM. (b) WECC CLM fitting results for ZIP+IM. (c) DDQN learning process for fitting CLOD's dynamics using WECC CLM. (d) WECC CLM fitting results for CLOD.

To further evaluate how the WECC CLM approximates other composite load models, we conduct a group of tests using WECC CLM to approximate the load dynamics generated from ZIP+IM and CLOD [38], [39] at bus 20 in the IEEE 39-bus system. The parameters for CLOD are selected following WECC 2001 CLOD generic parameters [38]. Then, we treat the dynamic responses from the ZIP+IM and CLOD load model as the field measurements, and our DDQN agent is trained to identify WECC CLM parameters to reproduce these P, Q dynamics. The training progress for fitting the dynamics from the ZIP+IM is shown in Fig. 21(a), and the training converges after 3,000 episodes. The P, Q fitting results are shown in Fig.

21(b), and the fitting RMSE of P and Q are shown in Table IIX. Similarly, the training results for CLOD is shown in Fig. 21(c), (d) and Table IIX. For CLOD, the training converges after 1,000 episodes.

According to the test result, our proposed method can accurately replicate the dynamics of a ZIP+IM using WECC CLM. However, due to the unique components contained in CLOD, only the P fitting accuracy using WECC CLM is satisfactory, and the Q fitting accuracy is not as good as the one in P.

TABLE IIX
FITTING RESULTS

| ZIP+IM | ZIP (%) 60 | | | | IM (%) 40 | |
|---|---|---|---|---|---|---|
| **Identified:** WECC CLM | IM_A | IM_B | IM_C | IM_D | ZIP | ELC |
| | 18.67% | 16.67% | 16.67% | 16.67% | 16.67% | 14.67% |
| | RMSE for P: | 0.0074 | | RMSE for Q: | 0.0161 | |

| CLOD | Small Motor | Large Motor | Discharging Lighting | Transformer Saturation | Constant MVA | Kp | R | X |
|---|---|---|---|---|---|---|---|---|
| | 10% | 10% | 0% | 0% | 0% | 1 | 0 | 0 |
| **Identified:** WECC CLM | IM_A | IM_B | IM_C | IM_D | ZIP | ELC | | |
| | 1.67% | 16.67% | 15.67% | 16.67% | 32.67% | 16.67% | | |
| | RMSE for P: | 0.0035 | | RMSE for Q: | 0.0611 | | | |

## V. CONCLUSIONS

In this paper, we develop a two-stage load modeling and identification method for WECC CLM. The first stage determines the load composition, and the second stage identifies the load parameters. This method offers the following contributions and advantages: it requires very limited prior knowledge towards hard-to-obtain and constantly updating load structure statistics. It is also scalable, from conventional composite load model such as ZIP + IM to complex load models such as the WECC CLM, or even more complex load models when additional load components are added. In addition, the identified load model using the proposed method is robust to different fault types and faults that occur at different locations. Furthermore, unlike common data-hungry methods that rely on a large number of disturbances data to calibrate, the proposed method only requires a set of reference dynamic responses, which is much more convenient to obtain. In the future, we will apply the proposed method to study more sophisticated load models, such as the distribution networks. In addition, we will also investigate the identification of the time-varying load model parameters considering the parameter sensitivity and dependency.

## REFERENCES

[1] P. Kundur, *Power System Stability and Control*. New York: McGrawHill, 1993.
[2] J. V. Milanovic, K. Yamashita, S. Martínez Villanueva, S. Ž. Djokic and L. M. Korunović, "International industry practice on power system load modeling," *IEEE Transactions on Power Systems*, vol. 28, no. 3, pp. 3038-3046, Aug. 2013.
[3] Y. Li, H.-D. Chiang, B.-K. Choi, Y.-T. Chen, D.-H. Huang, and M.G. Lauby, "Load models for modeling dynamic behaviors of reactive loads: Evaluation and comparison," *International Journal of Electrical Power & Energy Systems*, vol. 30, no. 9, pp. 497-503, 2008.
[4] M. Jin, H. Renmu, and D. J. Hill, "Load modeling by finding support vectors of load data from field measurements," *IEEE Transactions on Power Systems*, vol. 21, no. 2, pp. 726-735, May 2006.
[5] C. Wang, Z. Wang, J. Wang, and D. Zhao, "Robust time-varying parameter identification for composite load modelin," *IEEE Transactions on Smart Grid*, vol. 10, no. 1, pp. 967-979, Jan. 2019.
[6] D. Han, J. Ma, R. He et al., "A real application of measurement-based load modeling in large-scale power grids and its validation," *IEEE Transactions on Power Systems*, vol. 24, no. 4, pp. 1756-1764, Nov. 2009.
[7] P. Ju et al., "Composite load models based on field measurements and their applications in dynamic analysis," *IET Generation, Transmission & Distribution*, vol. 1, no. 5, pp. 724-730, Sep. 2007.
[8] A. Borden and B. Lesieutre, "Model validation: FIDVR event," Univ. of Wisconsin-Madison, Madison, WI, USA, Tech. Rep., 2009.
[9] D. Kosterev et al., "Load modeling in power system studies: WECC progress update," in *2008 IEEE Power and Energy Society General Meeting*, Pittsburgh, PA, 2008, pp. 1-8.
[10] Q. Huang, R. Huang, B. J. Palmer, Y. Liu, S. Jin, et al., "A generic modeling and development approach for WECC composite load model, " *Electric Power Systems Research*, vol. 172, pp 1-10, 2019.
[11] A. Gaikwad, P. Markham, and P. Pourbeik, "Implementation of the WECC composite load model for utilities using the component-based modeling approach," in *2016 IEEE/PES Transmission and Distribution Conference and Exposition (T&D)*, Dallas, TX, 2016, pp. 1-5.
[12] P. Etingov, "Load model data tool (LMDT)," https://svn.pnl.gov/LoadTool.
[13] J. Kim et al., "Fast and reliable estimation of composite load model parameters using analytical similarity of parameter sensitivity," *IEEE Transactions on Power Systems,* vol. 31, no. 1, pp. 663-671, Jan. 2016.
[14] K. Zhang, H. Zhu, and S. Guo, "Dependency analysis and improved parameter estimation for dynamic composite load modeling," *IEEE Transactions on Power Systems*, vol. 32, no. 4, pp. 3287-3297, Jul. 2017.
[15] S. Son et al., "Improvement of composite load modeling based on parameter sensitivity and dependency analyses," *IEEE Trans. Power Syst.*, vol. 29, no. 1, pp. 242–250, Jan. 2014.
[16] Y. Lin, Y. Wang, J. Wang, S. Wang, D. Shi, "Global Sensitivity Analysis in Load Modeling via Low-rank Tensor," IEEE Transactions on Smart Grid, 2020.
[17] Y. Li, Y. Wen, D. Tao and K. Guan, "Transforming Cooling Optimization for Green Data Center via Deep Reinforcement Learning," *IEEE Transactions on Cybernetics*. Doi: 10.1109/TCYB.2019.2927410.
[18] M. Schaarschmidt, F. Gessert, V. Dalibard, E. Yoneki, "Learning Runtime Parameters in Computer Systems with Delayed Experience Injection," in *2016 Conference on Neural Information Processing Systems (NIPS)*, Barcelona, Spain, Dec. 2016, pp. 1-10.
[19] A. Haj-Ali, N. K. Ahmed, T. Willke, J. Gonzalez, K. Asanovic, and I. Stoica, "A view on deep reinforcement learning in system optimization," arXiv:1908.01275, 2019.
[20] K. Zhang, S. Guo, and H. Zhu. "Parameter Sensitivity and Dependency Analysis for the WECC Dynamic Composite Load Model," *2017 50th Hawaii International Conference on System Sciences (HICSS)*, Waikoloa, HI, 2016, pp. 1-8.
[21] Western Electricity Coordinating Council, Technical Reference Document: "WECC Composite Load Model Specifications" Jan. 2015, https://www.wecc.org/
[22] S. Guo, T. J. Overbye, "Parameter estimation of a complex load model using phasor measurements," in *Proc. Power Energy Conf.*, Illinois, Feb. 2012, pp. 1–6.
[23] B. K. Choi, H. D. Chiang, "Multiple solutions and plateau phenomenon in measurement-based load model development: Issues and suggestions," *IEEE Trans. Power Syst.*, vol. 24, no. 2, pp. 824-831, May 2009.
[24] NERC, "Dynamic Load Modeling," Nov. 2016.
[25] Siemens Industry, Inc, software manual: "PSS®E 33.10, MODEL LIBRARY, " Apr. 2017.
[26] NERC Technical Reference Document: " A Look into load modeling: The composite load model." Sep. 2015, https://gig.lbl.gov/sites/all/files/6b-quint-composite-load-model-data.pdf
[27] Pacific Northwest National Laboratory, Technical Reference Document: "Composite load model evaluation," Sep. 2015.
[28] R. Koenker, K. Hallock, "Quantile regression: An introduction", *J. Econ. Perspectives*, vol. 15, no. 4, pp. 43-56, 2001.

[29] Q. Chang, Y. Wang, X. Lu et al., "Probabilistic Load Forecasting via Point Forecast Feature Integration," in *2019 IEEE Innovative Smart Grid Technologies - Asia (ISGT Asia)*, Chengdu, China, 2019, pp. 99-104.
[30] J. Duan, H. Xu and W. Liu, "Q-Learning-Based Damping Control of Wide-Area Power Systems Under Cyber Uncertainties," *IEEE Transactions on Smart Grid*, vol. 9, no. 6, pp. 6408-6418, Nov. 2018.
[31] J. Duan, Z. Yi, D. Shi, C. Lin, X. Lu and Z. Wang, "Reinforcement-Learning-Based Optimal Control of Hybrid Energy Storage Systems in Hybrid AC–DC Microgrids," *IEEE Transactions on Industrial Informatics*, vol. 15, no. 9, pp. 5355-5364, Sep. 2019.
[32] J. Duan, D. Shi et al., "Deep-Reinforcement-Learning-Based Autonomous Voltage Control for Power Grid Operations," *IEEE Transactions on Power Systems*, vol. 35, no. 1, pp. 814-817, Jan. 2020.
[33] H. V. Hasselt, A. Guez, and D. Silver, "Deep reinforcement learning with double Q-learning," arXiv:1509.06461,2016.
[34] I. Durugkar and P. Stone, "TD learning with constrained gradients," in *Proc. of the Deep Reinforcement Learning Symposium (NIPS 2017)*, Long Beach, CA, USA December 2017.
[35] R. Liu and J. Zou, "The effects of memory replay in reinforcement learning," *The ICML 2017 Workshop on Principled Approaches to Deep Learning*, Sydney, Australia, 2017.
[36] P. Cicilio and E. Cotilla-Sanchez, "Evaluating Measurement-Based Dynamic Load Modeling Techniques and Metrics," *IEEE Transactions on Power Systems, 2019*.
[37] R. J. Bravo, R. Yinger and P. Arons, "Fault Induced Delayed Voltage Recovery (FIDVR) Indicators," *2014 IEEE PES T&D Conference and Exposition*, Chicago, IL, 2014, pp. 1-5.
[38] A. S. Hoshyarzadeh, H. Zareipour, P. Keung, and S. S. Ahmed, "The Impact of CLOD Load Model Parameters on Dynamic Simulation of Large Power Systems," in *2019 IEEE International Conference on Environment and Electrical Engineering and 2019 IEEE Industrial and Commercial Power Systems Europe (EEEIC / I&CPS Europe)*, Genova, Italy, 2019, pp. 1-6.
[39] S. Li, X. Liang and W. Xu, "Dynamic load modeling for industrial facilities using template and PSS/E composite load model structure CLOD," in *2017 IEEE/IAS 53rd Industrial and Commercial Power Systems Technical Conference (I&CPS)*, Niagara Falls, ON, 2017, pp. 1-9.

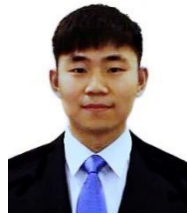
**Xinan Wang** (S'15) received the B.S. degree from Northwestern Polytechnical University, Xi'an, China, in 2013, and the M.S. degree from Arizona State University, Tempe, AZ, USA, in 2016, both in electrical engineering. He was a Research Assistant in the AI & System Analytics Group at GEIRI North America, San Jose, CA, in 2016, 2017 and 2019. He is currently pursuing the Ph.D. degree with the Department of Electrical and Computer Engineering at Southern Methodist University, Dallas, Texas, USA. His research interests include machine learning applications to power systems, wide-area measurement systems, data analysis and load modeling.

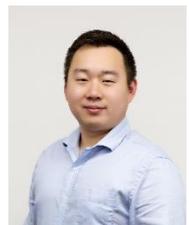
**Yishen Wang** (S'13–M'17) received the B.S. degree in electrical engineering from Tsinghua University, Beijing, China, in 2011, the M.S. and the Ph.D. degree in electrical engineering from the University of Washington, Seattle, WA, USA, in 2013 and 2017, respectively. He is currently a Power System Research Engineer with GEIRI North America, San Jose, CA, USA. His research interests include load modeling, PMU data analytics, power system economics and operation, energy storage, and microgrids.

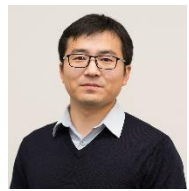
**Di Shi** (M'12–SM'17) received the Ph.D. degree in electrical engineering from Arizona State University. He is currently the Department Head of the AI & System Analytics Group, Global Energy Interconnection Research Institute North America (GEIRINA), San Jose, CA, USA. Prior to joining GEIRINA, He was a Research Staff Member with NEC Laboratories America. His research interests include PMU data analytics, AI, energy storage systems, the IoT for power systems, and renewable integration. He is an Editor of the IEEE Transactions on Smart Grid and the IEEE Power Engineering Letters.

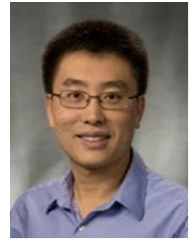
**Jianhui Wang** (M'07-SM'12). Dr. Jianhui Wang is an Associate Professor with the Department of Electrical and Computer Engineering at Southern Methodist University. Dr. Wang has authored and/or co-authored more than 300 journal and conference publications, which have been cited for more than 20,000 times by his peers with an H-index of 68. He has been invited to give tutorials and keynote speeches at major conferences including IEEE ISGT, IEEE SmartGridComm, IEEE SEGE, IEEE HPSC and IGEC-XI.

Dr. Wang is the past Editor-in-Chief of the IEEE Transactions on Smart Grid and an IEEE PES Distinguished Lecturer. He is also a guest editor of a Proceedings of the IEEE special issue on power grid resilience. He is the recipient of the IEEE PES Power System Operation Committee Prize Paper Award in 2015 and the 2018 Premium Award for Best Paper in IET Cyber-Physical Systems: Theory & Applications. Dr. Wang is a 2018 and 2019 Clarivate Analytics highly cited researcher for production of multiple highly cited papers that rank in the top 1% by citations for field and year in Web of Science.

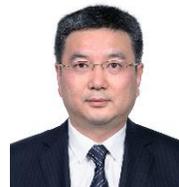
**Zhiwei Wang** (M'16-SM'18) received the B.S. and M.S. degrees in electrical engineering from Southeast University, Nanjing, China, in 1988 and 1991, respectively. He is President of GEIRI North America, San Jose, CA, USA. Prior to this assignment, he served as President of State Grid US Representative Office, New York City, from 2013 to 2015, and President of State Grid Wuxi Electric Power Supply Company from 2012-2013. His research interests include power system operation and control, relay protection, power system planning, and WAMS.